\documentclass{article}

\usepackage{arxiv}

\usepackage[utf8]{inputenc} 
\usepackage[T1]{fontenc}    
\usepackage{hyperref}       
\usepackage{url}            
\usepackage{booktabs}       
\usepackage{amsfonts}       
\usepackage{nicefrac}       
\usepackage{microtype}      
\usepackage{lipsum}

\usepackage{datetime,tabularx,multirow,subfigure,dblfloatfix}
\usepackage{graphicx,float}
\usepackage{fancyhdr,calc}
\usepackage{xcolor,colortbl,color}
\usepackage{soul,hyphenat}
\usepackage{algpseudocode}
\usepackage{algorithm}
\usepackage{textgreek, upgreek, tipa}
\usepackage{amssymb, amsthm, amsmath}
\usepackage{amssymb,amsthm}
\usepackage{relsize}
\usepackage{epstopdf}
\usepackage{siunitx}
\usepackage{booktabs}
\usepackage{balance}
\usepackage{enumitem}
\usepackage{placeins}
\usepackage{latexsym}
\usepackage{natbib}
\title{Learning to Invert Pseudo-Spectral Data for Seismic Waveform Inversion}

\author{
  Christopher Zerafa\\
  Department of Geosciences\\
  University of Malta\\
  Msida, Malta\\
  \texttt{christopher.zerafa.08@um.edu.mt} \\
   \And
   Pauline Galea \\
  Department of Geosciences\\
  University of Malta\\
  Msida, Malta\\
  \texttt{pauline.galea@um.edu.mt} \\
   \AND
   Cristiana Sebu \\
  Department of Geosciences\\
  University of Malta\\
  Msida, Malta\\
   \texttt{cristiana.sebu@um.edu.mt} 
}

\begin{document}
\maketitle

\begin{abstract}
Full-waveform inversion (FWI) is a widely used technique in seismic processing to produce high resolution Earth models that fully explain the recorded seismic data. FWI is a local optimisation problem which aims to minimise in a least-squares sense the misfit between recorded and modelled data. The inversion process begins with a best-guess initial model which is iteratively improved using a sequence of linearised local inversions to solve a fully non-linear problem. Deep learning has gained widespread popularity in the new millennium. At the core of these tools are Neural Networks (NN), in particular Deep Neural Networks (DNN) are variants of these original NN algorithms with significantly more hidden layers, resulting in efficient learning of a non-linear functional between input and output pairs. The learning process within DNN involves iteratively updating network neuron weights to best approximate input-to-output mappings. There is clearly similarity between FWI and DNN. Both approaches attempt to solve for a non-linear mapping in an iterative sense, however they are fundamentally different in that the former is knowledge-driven whereas the latter is data-driven. This article proposes a novel approach which learns pseudo-spectral data-driven FWI. We test this methodology by training a DNN on 1D multi-layer, horizontally-isotropic data and then apply this to previously unseen data to infer the surface velocity. Results are compared against a synthetic model and successfulness and failures of this approach are hence identified.
\end{abstract}


\section{Introduction}
\subsection{Preliminaries}
The seismic reflection method uses artificially generated seismic waves that excite the Earth and propagate through the subsurface. They are attenuated by interactions with their medium of propagation and are partially reflected back across a high contrasting acoustic impedance layer. A simple 2D two-layer example of an acoustic forward propagation through the subsurface is given in Figure \ref{fig:simple_2d_model}. The model contains a high acoustic impedance layer between 1 and 1.5km depth. When hitting the interface between different velocity layers, the wave is reflected back to the surface and recorded by receivers (geophones or hydrophones) located at or close to the surface. The internal structure of the subsurface can be then inferred from the recorded wave total travel time. 

\newcommand{\w}{0.3}
\begin{figure*}[!t]
	\centering
	\subfigure[Synthetic p-wave velocity model.]{\includegraphics[width=0.32\textwidth]{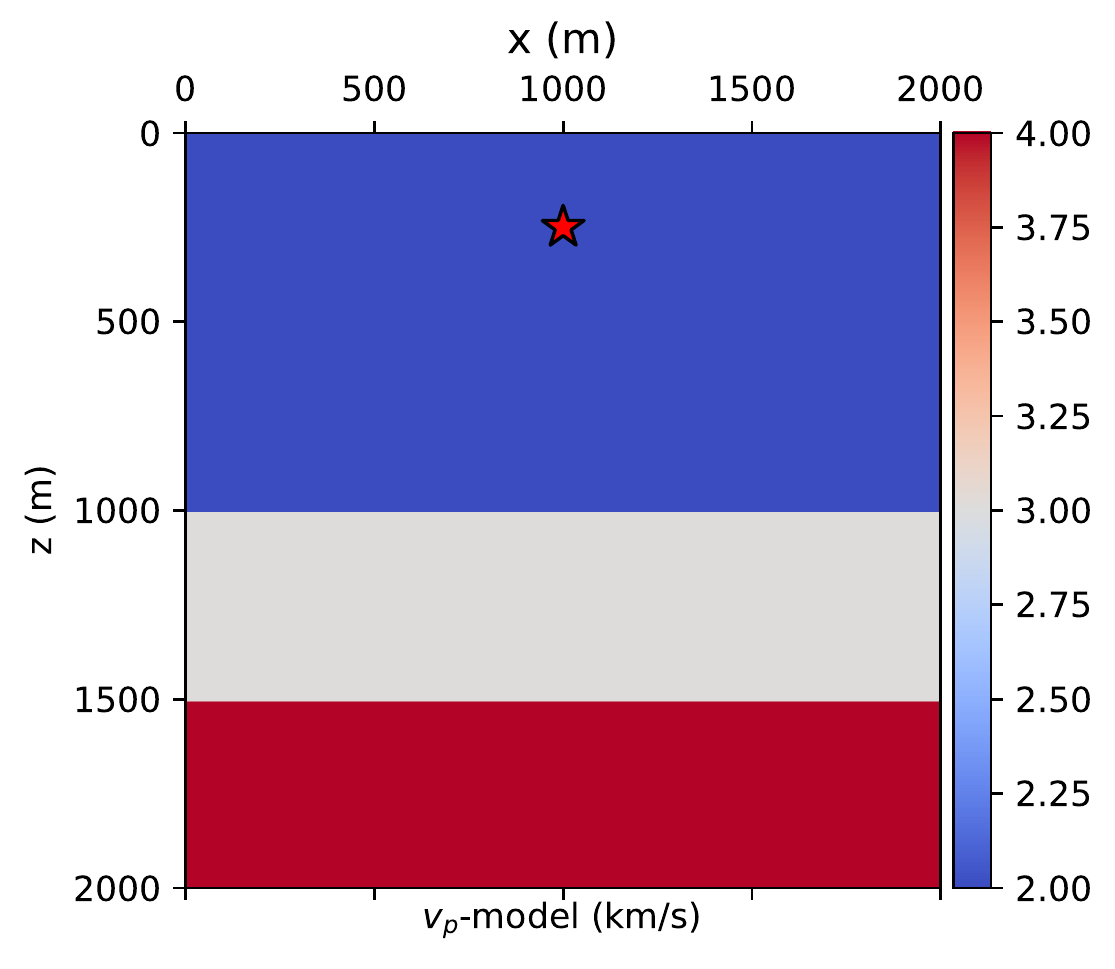}}
	\subfigure[Time-step 300ms.]{
		\includegraphics[width=\w\textwidth]{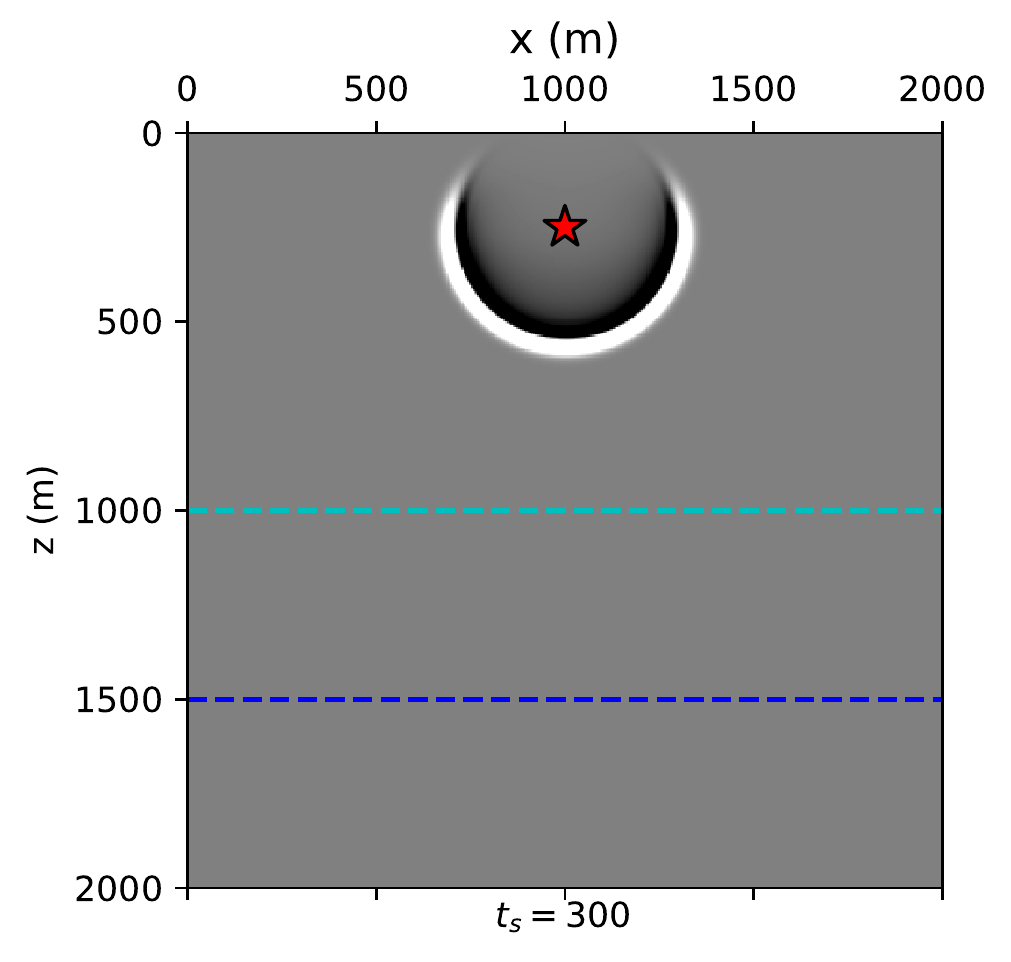}}
		\subfigure[Time-step 600ms.]{
			\includegraphics[width=\w\textwidth]{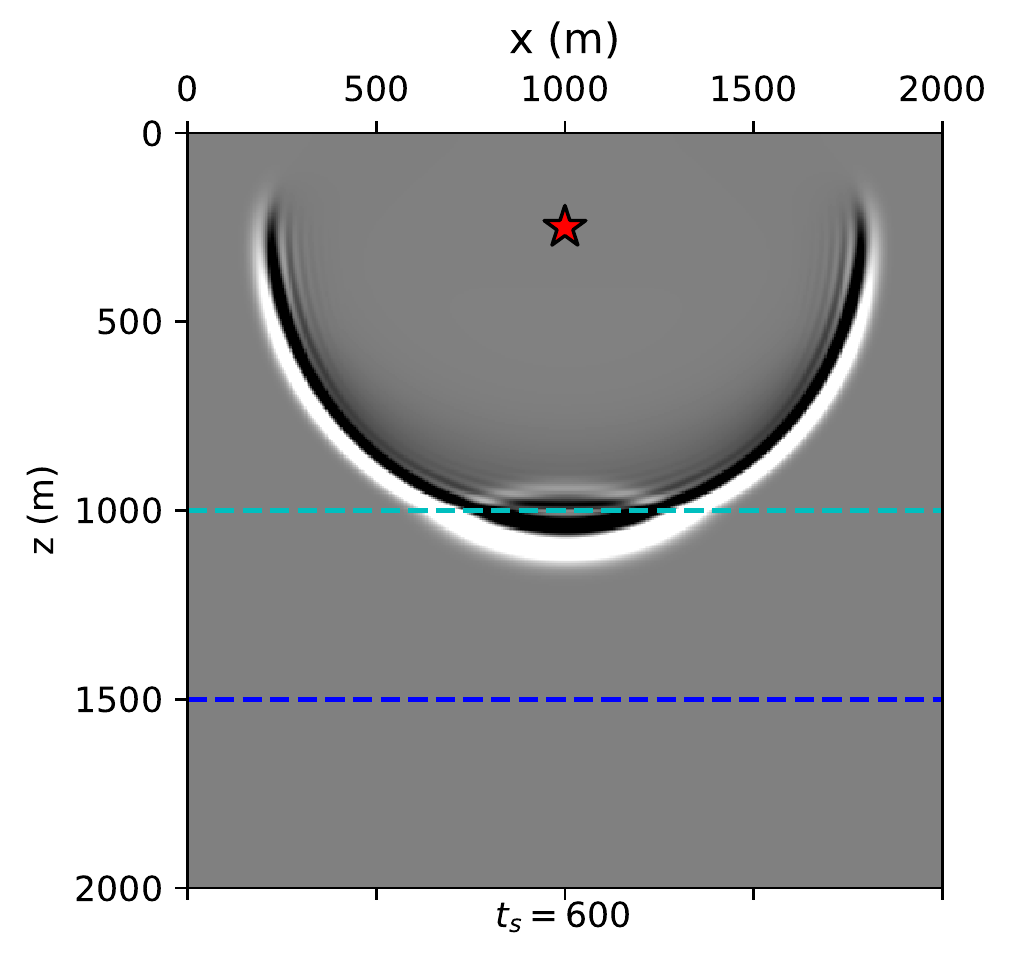}}
	\hfill
	\subfigure[Time-step 700ms.]{
		\includegraphics[width=\w\textwidth]{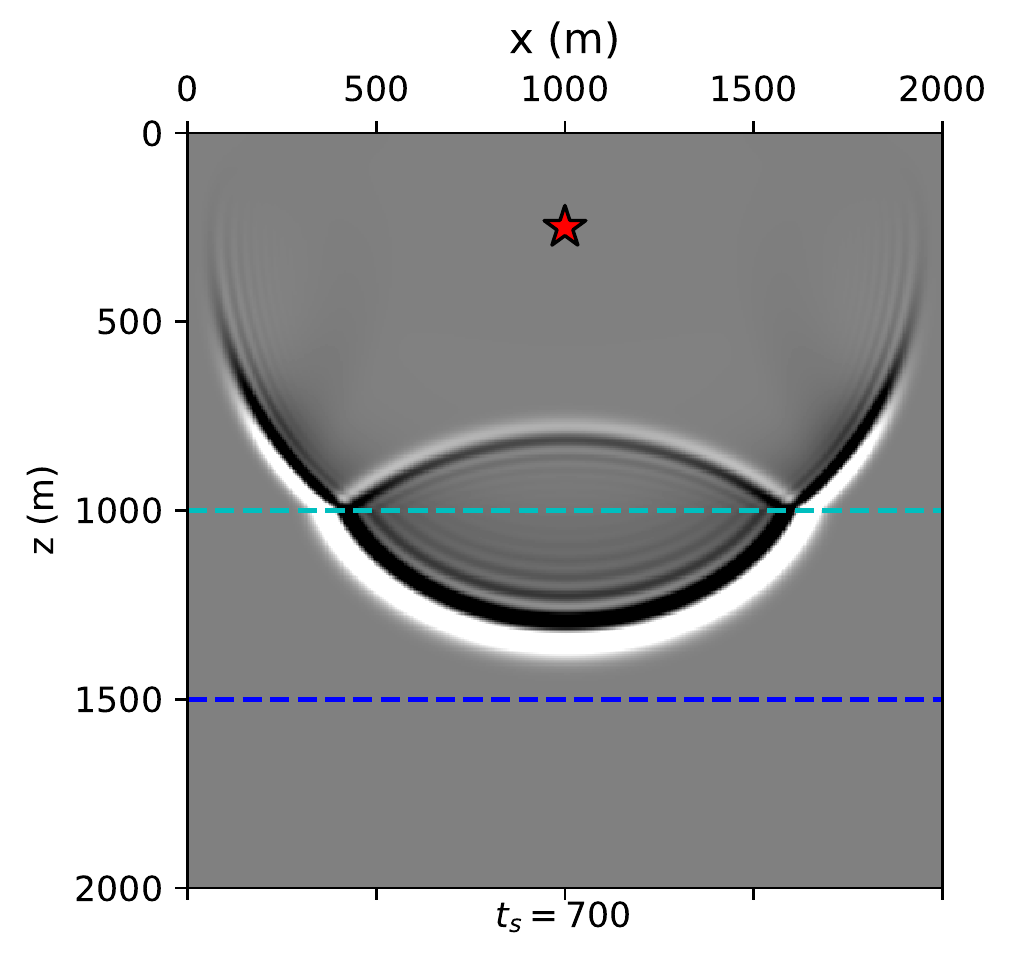}}
	\subfigure[Time-step 800ms.]{	
		\includegraphics[width=\w\textwidth]{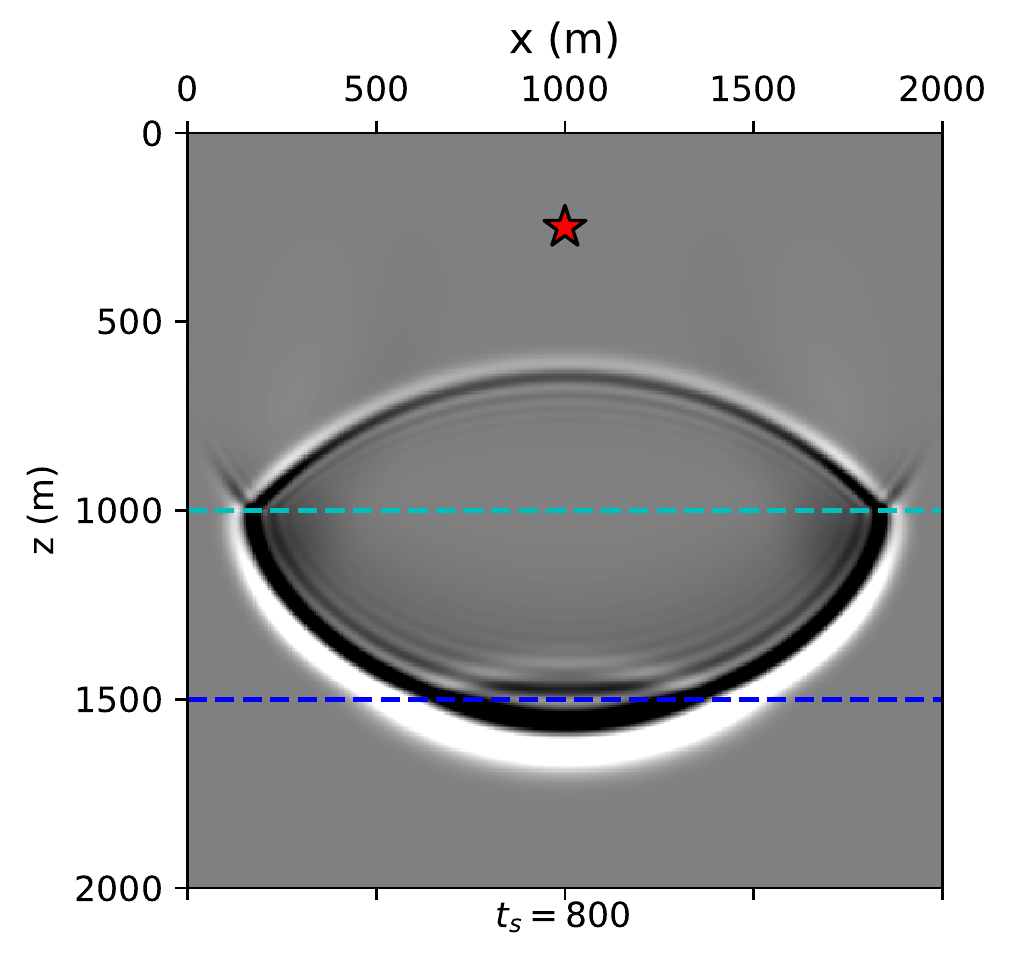}}
	\subfigure[Time-step 950ms.]{	
		\includegraphics[width=\w\textwidth]{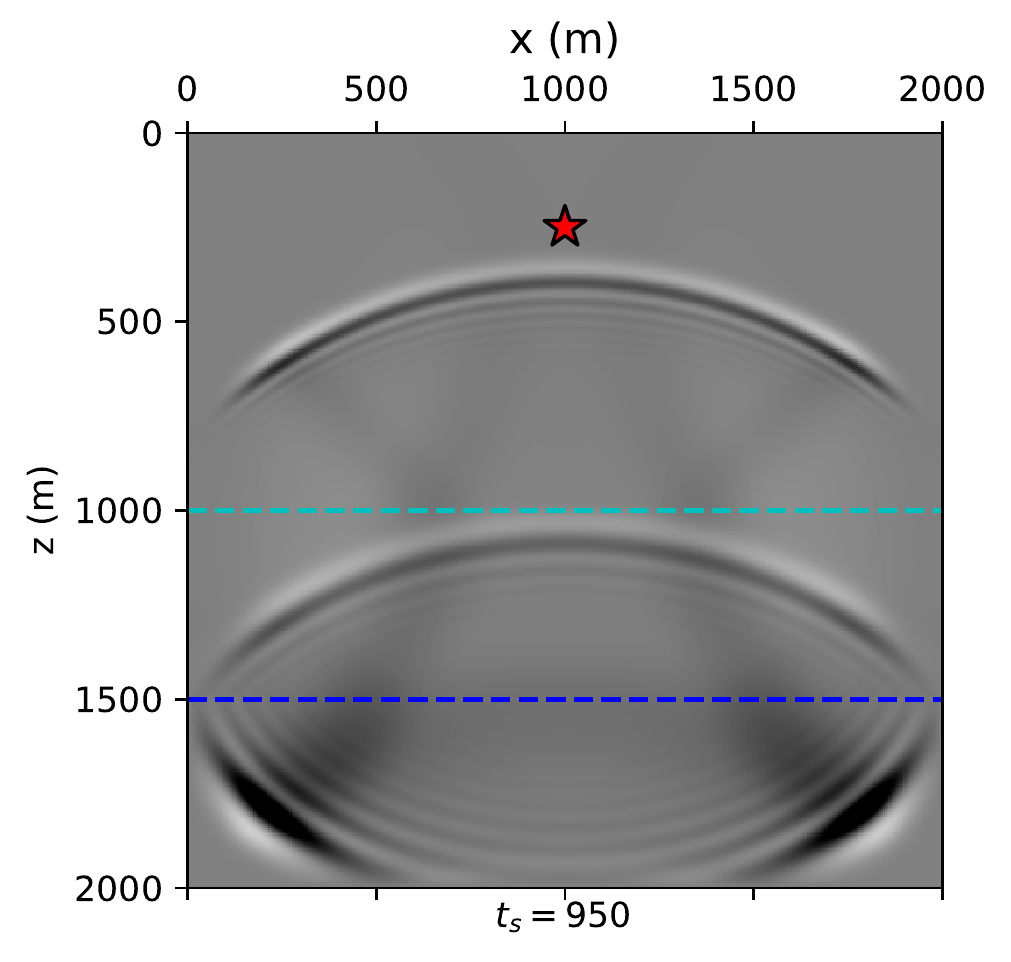}}
	\caption{Simple 2D two-layer model used for forward propagation of seismic waves. The red star marks the source location at time-step 0ms. Figure (a) is the ground truth velocity. Figures (b) to (f) illustrate the propagation of an acoustic wave through (a).}
	\label{fig:simple_2d_model}
\end{figure*}

\begin{figure*}[!t]
	\centering
	\subfigure[Conventional method.]{\label{fig:fwi_slice_0}
		\includegraphics[width=60mm]{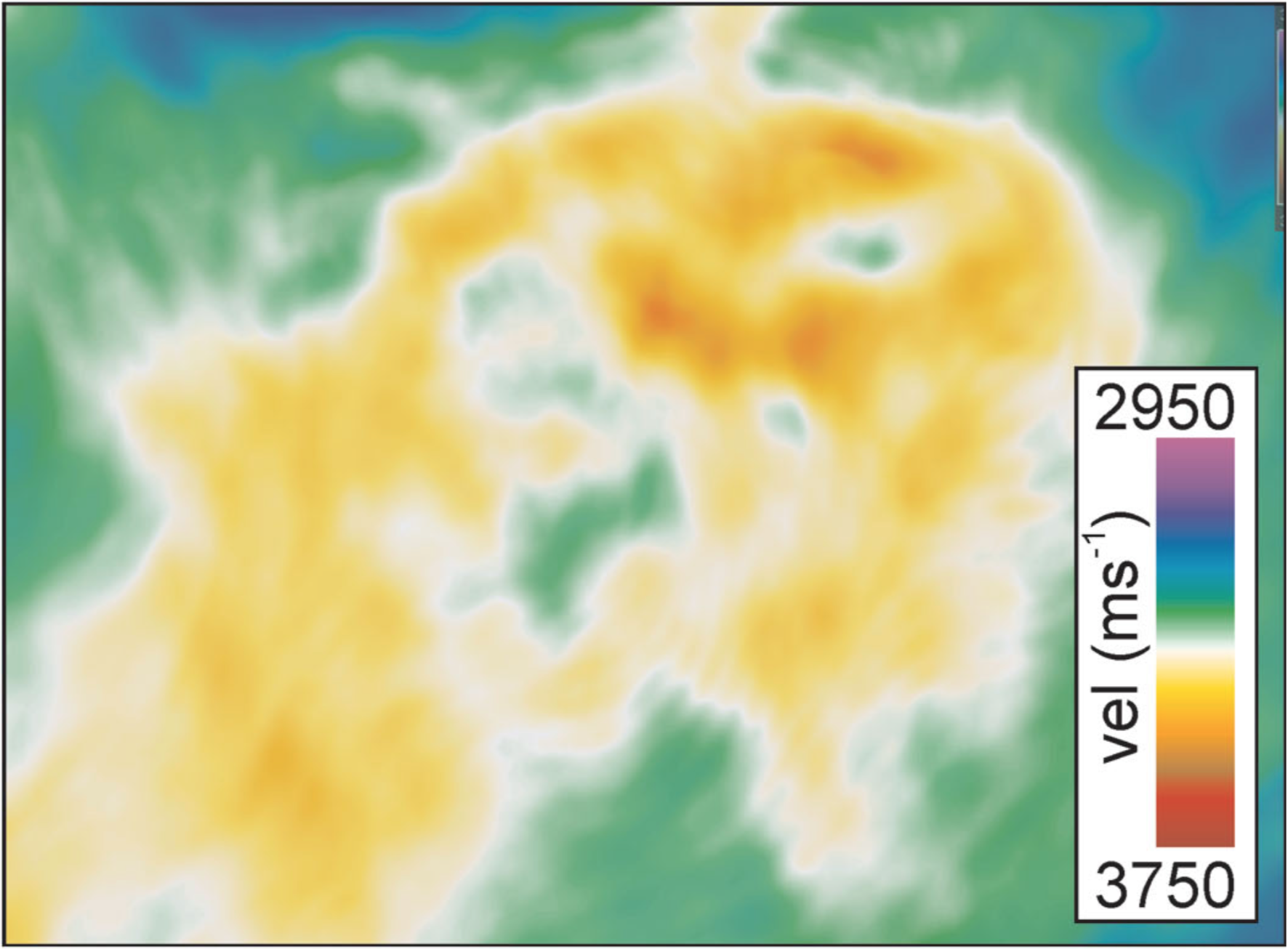}
	}
	\subfigure[ $<$10Hz FWI velocity model result.]{\label{fig:fwi_slice_1}
		\includegraphics[width=60mm]{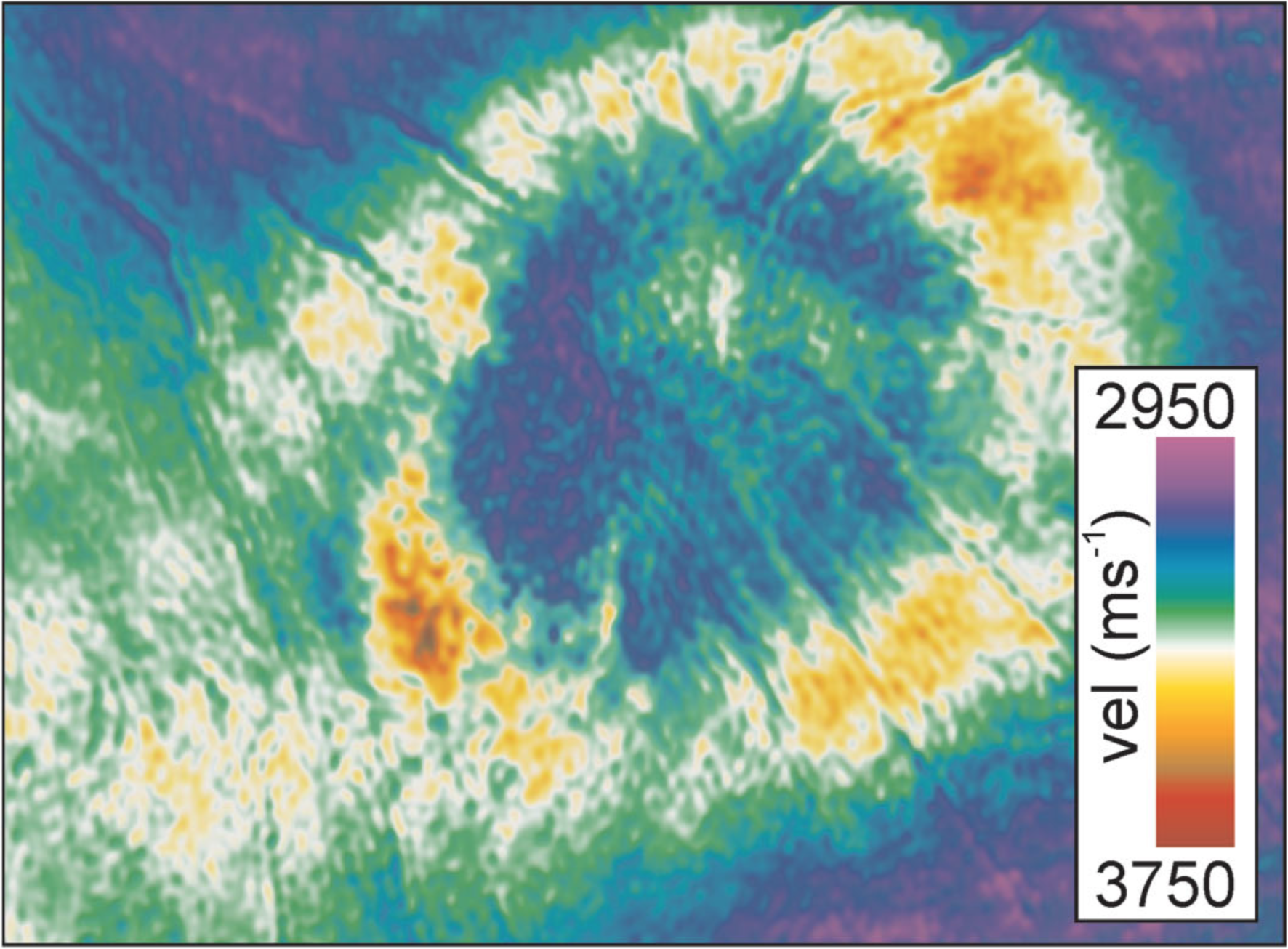}
	}
	\caption{Horizontal slices though the Samson Dome at 1350m. From \cite{morgan2013next}.}
	\label{fig:fwi_slices_time}
\end{figure*}

\begin{figure*}[!t]
	\centering
	\subfigure[Original 2004 BP synthetic for FWI.]{\label{fig:fwi_slice_2}
		\includegraphics[width=0.6\textwidth]{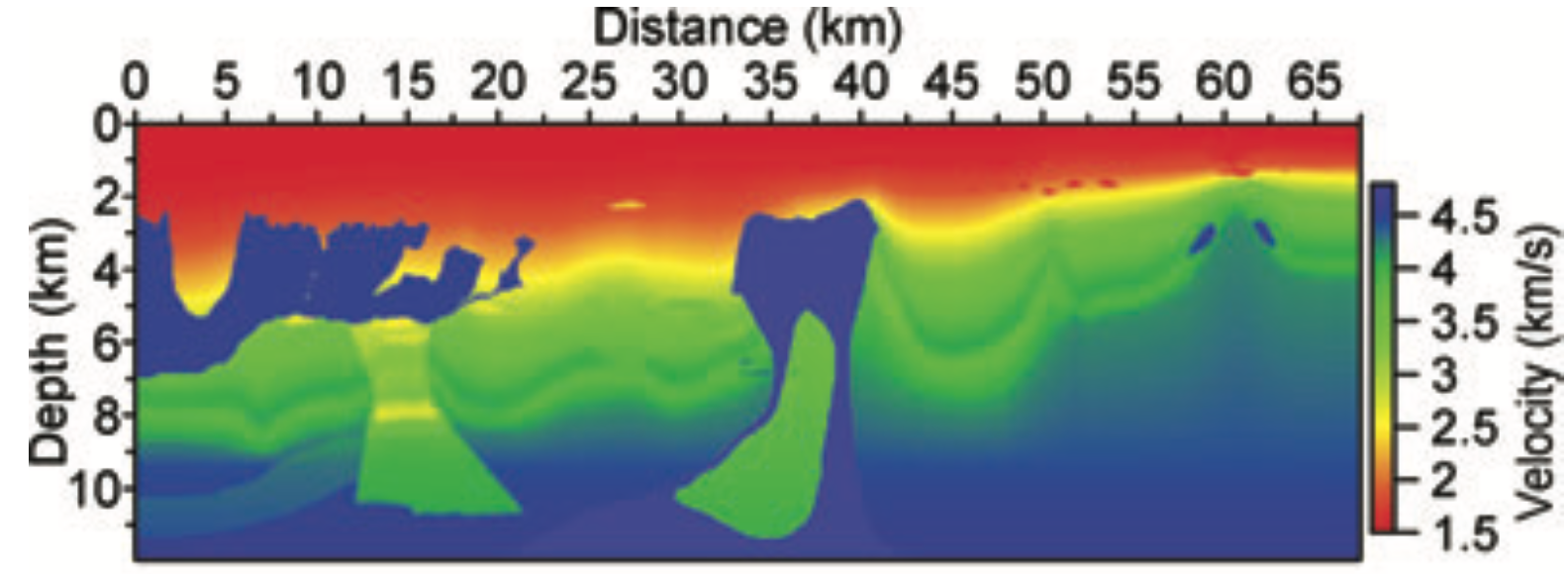}
	}
	\subfigure[Zoomed section.]{\label{fig:fwi_slice_2_zoom}
		\includegraphics[width=0.3\textwidth, trim = {24cm 4cm 22cm 9cm}, clip]{fwi_slice_2_synthetic.png}
	}
\hfill
	\subfigure[2D FWI result.]{\label{fig:fwi_slice_3}
		\includegraphics[width=0.6\textwidth]{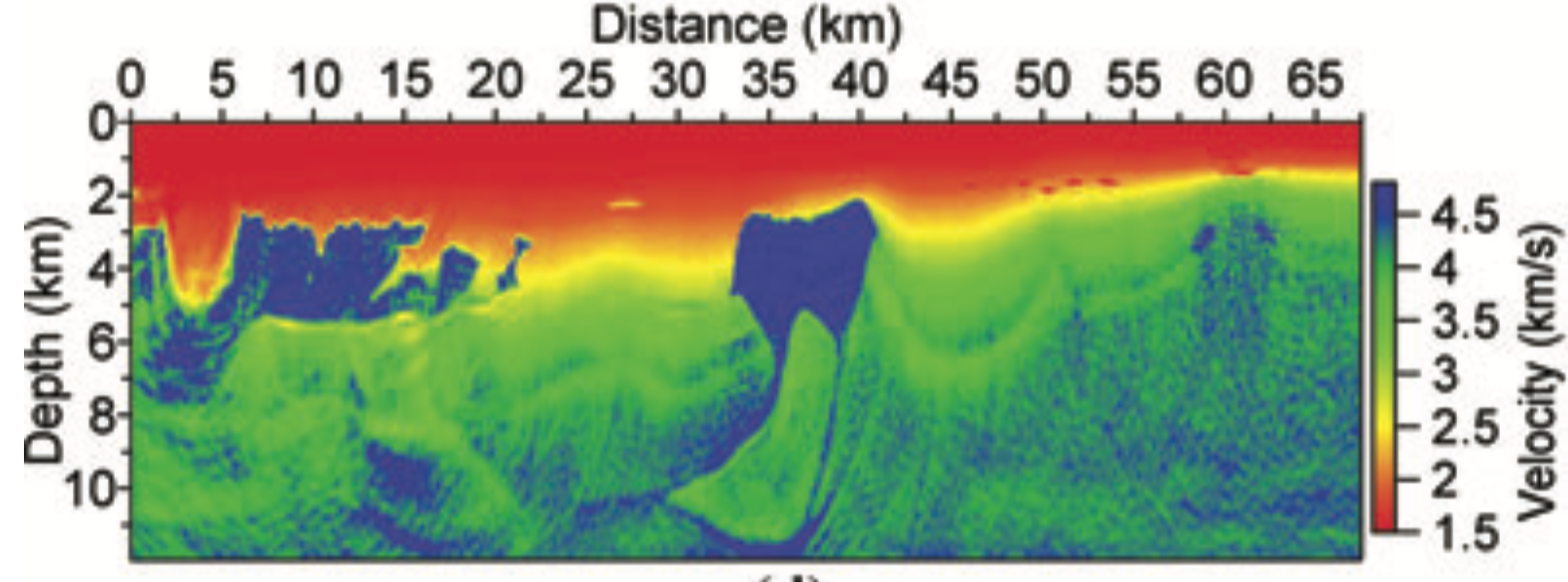}
	}
	\subfigure[Lack of resolution.]{\label{fig:fwi_slice_3_zoom}
		\includegraphics[width=0.3\textwidth, trim = {24cm 4cm 22cm 9cm}, clip]{fwi_slice_3_fwi_result.png}
	}
	\caption{Limitations of FWI in complicated geology. From \cite{shin2010sequentially}.}
	\label{fig:fwi_slices_depth}
\end{figure*}

Full-waveform inversion (FWI) is a technique which tries to exploit the information contained in the reflected seismic wave-field as much as possible. It goes beyond refraction and reflection tomography techniques which use only the travel time kinematics of the seismic data. It honours the Physics of the finite-frequency wave equation and uses the additional information provided by the amplitude and phase of the seismic waveform \cite{tarantola1987inverse}. FWI seeks to achieve a high-resolution geological model of the subsurface through application of multivariate optimisation to the seismic inverse problem \cite{lailly1983seismic, tarantola1984inversion, virieux2009overview}. The inversion process begins with a best-guess initial model which is iteratively improved using a sequence of linearised local inversions to solve a fully non-linear problem. Figure \ref{fig:fwi_slices_time} illustrates the imaging uplift which is achievable through FWI. In situations of more complex structures such as complicated salt structures with convoluted ray-paths in the overburden, the inversion becomes more difficult and computationally more expensive. Figure \ref{fig:fwi_slices_depth} illustrates an example of FWI on the 2004 BP synthetic data. The zoomed sections in Figure \ref{fig:fwi_slice_3_zoom} clearly illustrate a lack of resolution of FWI.

\subsection{Aims \& Objectives}
Optimization theory is fundamental to FWI. The parameters of the system under investigation are reconstructed from indirect observations that are subject to a forward modelling process \cite{tarantola2005inverse}. The accuracy of this forward modelling depends on the validity of physical theory that links ground-truth to the measured data \cite{innanenquantifying}. Moreover, solving for this inverse problem involves learning the inverse mapping from measurements to the ground-truth which is based on a subset of degraded best-estimated data \cite{tarantola2005inverse,Tikhonov1977}. Two limitations within inverse theory can be identified: (i) solving the forward problem and (ii) training the data. 

Choice of the numerical method used to solve the forward problem will crucially impact the accuracy of the FWI result. Challenging environments require more complex assumptions to explain the physical link between data and observations, with not necessarily improved levels of accuracies \cite{morgan2013next}. Secondly, the data being used to reconstruct the mapping of measurements for the ground-truth are not optimal. Very wide angle and multi-azimuth data are required to enable full inversion \cite{Morgan2016}; this information might not necessarily have been recorded in the acquisition stages of the data. Furthermore, pre-conditioning of data is a necessity prior to FWI to induce well-posedness \cite{Kumar2012a, Mothi2013, Peng2018, Warner2013}. However, if done incorrectly this can degrade the inversion process \cite{Lines2014}. Indeed, \cite{Lines2014} shows how FWI remains robust to both random and coherent noise, and his work indicates that FWI with the inclusion of multiple data proves useful at estimating a better solution in some situations.

Recently, deep learning (DL) techniques have emerged as excellent models and gained great popularity for their widespread success in pattern recognition \cite{Ciresan2012, Ciresan2011}, speech recognition \cite{Hinton2012} and computer vision \cite{Krizhevsky2015, Deng2013}. The use of Deep Neural Networks (DNN) to solve inverse problems has been explored by \cite{Elshafiey1991, Adler2017a, Chang2017, Wei2017} and has achieved state-of-the-art performance in image reconstruction \cite{Kelly2017, Petersen2017, Adler2017b}, super-resolution \cite{Bruna2015, Galliani2017} and automatic-colorization \cite{Larsson2016}. 

In Geophysics, the applications of DL techniques have focused on the identification of features and attributes in migrated seismic sections, with few studies looking into velocity inversion. \cite{zhang2014machine} used a kernel regularized least-squares method for fault detection from seismic records on numerical experiments. \cite{wang2018automatic} employed a fully convolutional neural network (FCN) to perform salt-detection from raw multi-shot gathers which was found to be much faster and efficient than traditional migration and interpretation. \cite{lewis2017deep} combined DL and FWI to improve the performance for salt inversion by generating a probability map from learned abstractions of the data and incorporating these in the FWI objective function. These tests results showed promise for automated salt body reconstruction using FWI. \cite{mosser2018rapid} used a generative adversarial network \cite{Goodfellow-et-al-2016} with cycle-constraints to perform seismic inversion by reformulating the inversion problem as a domain-transfer problem. The mapping between post-stack seismic traces and p-wave velocity models was approximated through DL. More recently, \cite{yang2019deep} developed a supervised FCN for velocity-model building directly from raw seismograms using a DNN architecture based on U-Net \cite{ronneberger2015u}. Their training data was obtained from modelling of the acoustic wave equation via a time-domain staggered-grid finite-difference scheme, with numerical experiments showing good potential of DL for seismic velocity inversion.

In this work, we are re-casting the mathematical formulation of FWI within a DL framework. The conventional least-squares formulation of FWI can be expressed as:
\begin{equation}
\min_\mathbf{m} \emph{J}(\mathbf{m}) = {\left|| \mathbf{d} - F(\mathbf{m}) \right||}_2^2,
\end{equation}
where $\mathbf{m}\in M$ is the subsurface model, $F:M\rightarrow D$ is the forward wave equation model, and $\mathbf{d}\in D$ is the observed data. This inversion is non-linear and ill-posed since $\mathbf{d}$ does not contain all subsurface information to define a velocity model explicitly \cite{Biondi2006}. Based on the Universal Approximation Theorem \cite{Hornik1989}, a DNN can be used to approximate the non-linear inverse operator $F^{-1}:D\rightarrow M$ by a pseudo-inverse operator or mapping function $g_{\theta}$ which minimizes the functional:
\begin{equation}
\emph{J}(\mathbf{m}) = {\left|| \mathbf{m} - g_{\theta}(\mathbf{d})\right||}^2,
\end{equation}
where $\theta$ is a large simulated dataset of pairs $(\mathbf{m}, \mathbf{d})$ used for learning the process function $g_{\theta}$ \cite{Hastie2001}. In particular, based on the work of \cite{Falsaperla1996}, DNN utilizing pseudo-spectral transformed data $\mathcal{F}$ facilitates the learning process due to better sparsity in the transformed domain, as compared to the time domain. The novelty of this approach is the combination of both DL, signal processing and inverse theory for subsurface velocity inversion. This papers aims to prove what theory indicates is a potentially viable solution via a practical implementation to a 1D synthetic model.

The structure of this manuscript is as follows. Section 1 introduces the subject of FWI and its importance within current workflows for seismic exploration. Limitations within the current formulation are identified and a novel approach to devise better velocity models of the subsurface is proposed. In Section 2,  mathematical fundamentals for FWI and DNN are derived respectively. These are then compared and their differences are highlighted. In particular, FWI is recast as a DL problem. Based on the derived formulation in Section 2, numerical results of this novel approach are presented in Section 3 and a 1D synthetic highlights the successfulness and failures of this approach. In Section 4, concluding remarks are presented.

\section{Theoretical framework and methodology}
\subsection{Inverse problem formulation}
The aim of inversion is to estimate the parameters of a physical system based on the measurements available. In the case of Geophysics, the physical system is the Earth and data are the recorded wave-field. 
\begin{figure}[h!]
	\centering
	\includegraphics[width=0.45\textwidth, 
	trim = {3cm 20.5cm 5.8cm 2cm}, clip]{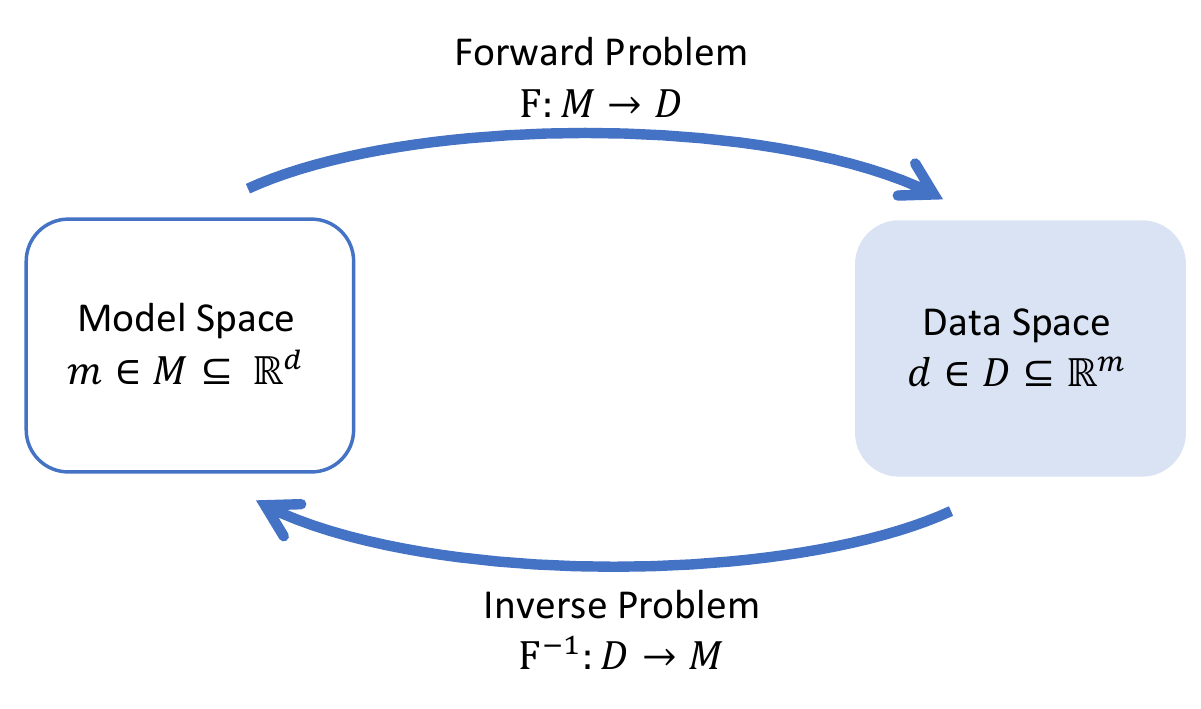}
	\caption{Visual representation of the mapping between the Forward and the Inverse problem.}
	\label{fig:forw_inv}
\end{figure}

The recorded wave-field is known, while the physical properties of the medium which the wave-field propagated through are the unknowns. The wave-field will be a function of these medium properties and the function for the forward problem can be as expressed as:
\begin{equation}
\mathbf{d}=F(\mathbf{m}),
\end{equation}
where $F:M\rightarrow D, F\in\mathbb{R}^{(m\times d)}$ is the operator applied on the model space $\mathbf{m}\in M \subseteq \mathbb{R}^d $ to recover measurements $\mathbf{d} \in D \subseteq \mathbb{R}^m$. The forward problem is well-posed, that is, a unique solution exists that depends continuously on the model in some appropriate topology. 

The opposite to forward modelling is the inversion. This involves making assumptions on the physical properties of the object we want to image to be able to compute the wave-field at any given time and location to a certain degree of accuracy. If $F$ is invertible, the inverse problem is given by:
\begin{equation}
\mathbf{m} = F^{-1}(\mathbf{d}),
\end{equation}
This aims to extract all the information contained within the data. 

\subsection{FWI as local optimisation}
\cite{Lailly1983} and \cite{tarantola1984inversion} re-cast the migration imaging principle introduced by \cite{Claerbout1971} as a local optimisation problem. The forward problem is based on the wave equation, which is one of the most fundamental equations in Physics used for the description of wave motion. It is a second order, partial differential equation involving both time and space derivatives.

The particle motion for an isotropic medium is given by:
\begin{equation}
\frac{1}{c(\mathbf{m})^2} \frac{\partial^2p(\mathbf{m},t)}{\partial t^2} - \nabla^2p(\mathbf{m},t) = s(\mathbf{m},t),
\end{equation}
where $p(\mathbf{m},t)$ is the pressure wave-field, $c(\mathbf{m})$ is the acoustic $p$-wave velocity and $s(\mathbf{m},t)$ is the source \cite{Igel2016}. To solve the wave equation numerically, it can be expressed as a linear operator. Although the data $\mathbf{d}$ and model $\mathbf{m}$ are not linearly related, the wave-field $p(\mathbf{m},t)$ and the sources $s(\mathbf{m},t)$ are linearly related by the equation:
\begin{equation}
\mathbf{A}p(\mathbf{m},t) = s(\mathbf{m},t),
\end{equation}
where $p(\mathbf{m},t)$ is the pressure wave-field produced by a source $s(\mathbf{m})$ and $\mathbf{A}$ is the numerical implementation of the operator:
\begin{equation}
\frac{1}{c(\mathbf{m})^2} \frac{\partial^2}{\partial t^2} - \nabla^2,
\end{equation}

A common technique employed within the forward modelling stage is to perform modelling in pseudo-spectral domain ($\mathcal{F}$) rather than the time domain ($\mathcal{T}$). The most common domain is the Fourier domain \cite{Igel2016} and computational implementation is generally achieved via the Fast Fourier Transform (FFT) developed by \cite{cooley1965algorithm} as it utilises the fact that $e^{-2\pi i/N}$ is an $N$-th primitive root of unity and allows for the reduction of computational costs from $O(N^2)$ to $O(N\log N)$. 

After forward modelling the data in pseudo-spectral domain, the objective is to seek to minimize the difference between the observed data and the modelled data. The metric for the difference or misfit between the two datasets is known as the misfit-, objective- or cost-function $\emph{J}$. The most common cost function is given by the $L_2$-norm of the data residuals:
\begin{equation}
\emph{J}(\mathbf{m}) = \frac{1}{2}{\left|| d - F(\mathbf{m}) \right||}^2_D,
\end{equation}
where $D$ indicates the data domain given by $n_s$ sources and $n_r$ receivers \cite{Igel2016}. The misfit function $\emph{J}$ can be minimized with respect to the model parameters $d$ if the gradient is zero, namely:
\begin{equation}
\nabla\emph{J} = \frac{\partial\emph{J}}{\partial \mathbf{d}} = 0,
\end{equation}

Minimising the misfit function is generally achieved via a linearised iterative optimisation scheme based on the Born approximation in scattering theory \cite{Born1980, Clayton1980}. The inversion algorithm starts with an initial estimate of the model $\mathbf{m}_0$. After the first pass via forward modelling, the model is updated by the model parameter perturbation $\Delta \mathbf{m}_0$. This newly updated model is then used to calculate the next update and the procedure continues iteratively until the computed model is close enough to the observations based on a residual threshold criterion. At each iteration $k$, the misfit function $\emph{J}(\mathbf{m}_k)$ is calculated from the previous iteration model $\mathbf{m}_{k-1}$ by:
\begin{equation} \label{eq:misfit_k-1}
\emph{J}(\mathbf{m}_k) = \emph{J}(\mathbf{m}_{k-1} + \Delta \mathbf{m}_0),
\end{equation}
Assuming that the model perturbation is small enough with respect to the model, equation (\ref{eq:misfit_k-1}) can be expanded via Taylor series up to second orders as:
\begin{equation} \label{eq:tay_exp}
	\begin{split}
		\emph{J}(\mathbf{m}_k) &= \emph{J}(\mathbf{m}_{k-1} + \Delta \mathbf{m}_0)\\
		& = \emph{J}(\mathbf{m}_{k-1}) 
		+ \delta \mathbf{m}^T_{k-1}
		\frac{\partial\emph{J}}{\partial\mathbf{m}_{k-1}}
		+ \frac{1}{2}
		\delta\mathbf{m}^{2T}_{k-1}
		\frac{\partial^2\emph{J}}{\partial\mathbf{m}^2_{k-1}},
\end{split}
\end{equation}
Taking the derivative of equation (\ref{eq:tay_exp}) and minimizing to determine the model update leads to:
\begin{equation}
\delta\mathbf{m}_{k-1} \approx -\mathbf{H}^{-1}\nabla_{\mathbf{m}_{k-1}}\emph{J},
\end{equation}
where $\mathbf{H} = \frac{\partial^2\emph{J}}{\partial\mathbf{m}^2_{k-1}}$ is the Hessian matrix and $\nabla_{\mathbf{m}_{k-1}}\emph{J}$ the gradient of the misfit function. The Hessian matrix is a symmetric matrix of size $N\times N$ where $N$ is the number of model parameters and represents the curvature trend of the quadratic misfit function.

FWI is an ill-posed problem, implying there exist an infinite number of models that fit the observations. Well-posedness can be introduced with the addition of Tikhonov $L_2$-norm regularization \cite{Tikhonov1963, Tikhonov1977}:
\begin{equation}
\emph{J}(\mathbf{m}) = \frac{1}{2}\left[ {\left|| d - F(\mathbf{m}) \right||}^2_D + \lambda\left||\mathbf{m}\right||^2_M \right] ,
\end{equation}
where $\lambda$ is the regularization parameter which signifies the trade-off between the data and model residuals. 

\subsection{FWI algorithm summary}
\begin{figure*}[b!]
	\centering
	\includegraphics[width=0.75\textwidth, trim = {1.5cm 1cm 1.5cm 2.5cm}, clip]{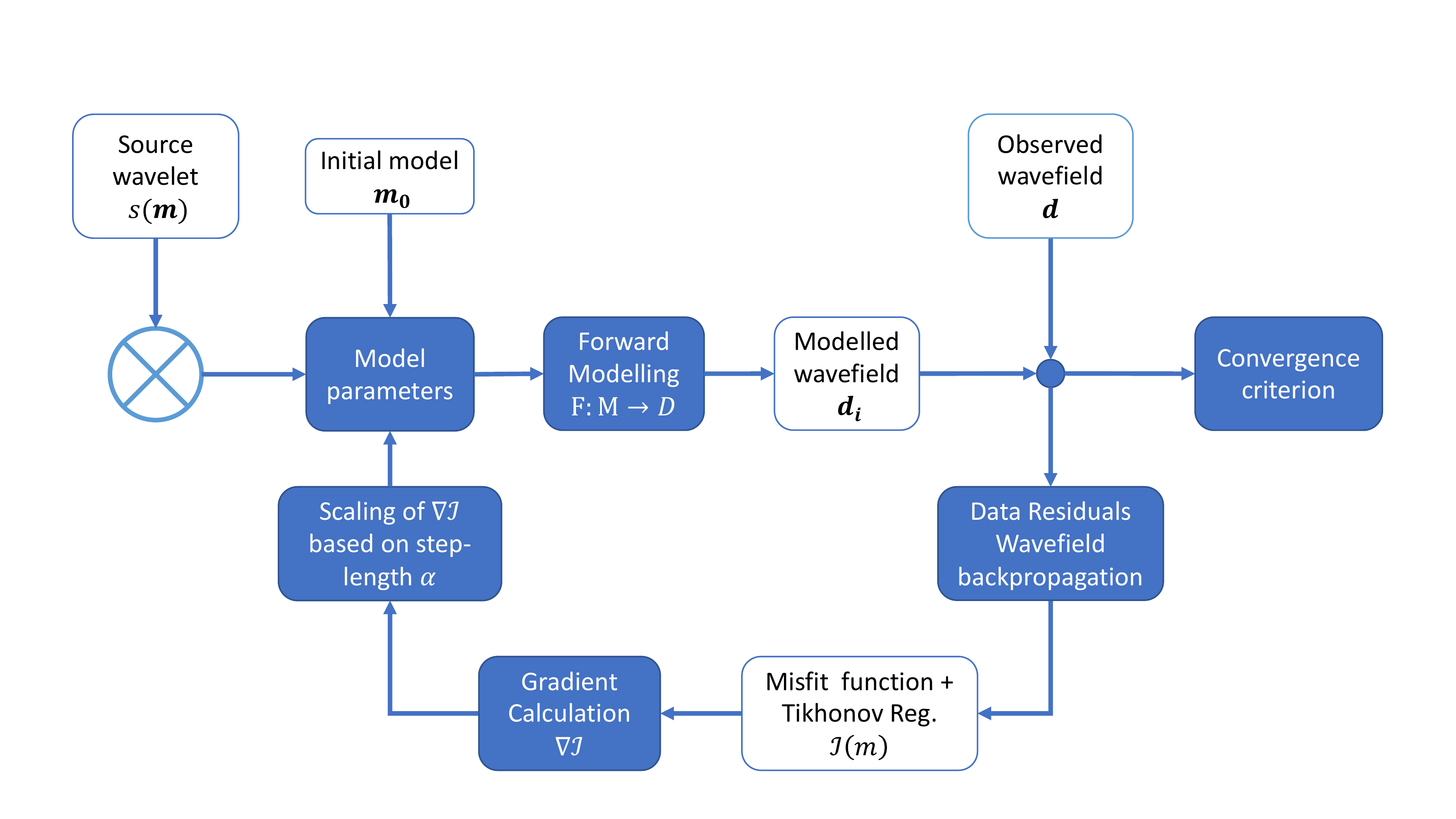}
	\caption{Schematic of a FWI workflow solved as an iterative optimisation process.}
	\label{fig:algo_schem_fwi}
\end{figure*}

A summary of FWI as a local optimisation problem is given in Algorithm \ref{algo:fwi} and a schematic is illustrated in Figure \ref{fig:algo_schem_fwi}.

\begin{algorithm}
	\caption{FWI as a local optimisation problem}
	\renewcommand{\labelenumi}{(\Roman{enumi})}
	\begin{enumerate}[noitemsep,nolistsep]
		\item Choose an initial model $\mathbf{m}_0$ and source wavelet $s(\mathbf{m})$.
		\item \label{itm:algo_fwi_2}  For each source location, solve the forward problem $F:M\rightarrow D$ using pseudo-spectral forward modelling everywhere in the model space to get a predicted wave-field $\mathbf{d}_k$. This is sampled at receivers $r(\mathbf{m})$.
		\item At every receiver $r(\mathbf{m})$, data residuals are calculated between the modelled wave-field $\mathbf{d}_k$ and the observed data $\mathbf{d}$.
		\item These data residuals are back-propagated from the receivers to produce a back-propagated residual wave-field.
		\item For each source location, the misfit function $\emph{J}(\mathbf{m})$ is calculated for the observed data and back-propagated residual wave-field to generate the gradient $\nabla\emph{J}$ required at every point in the model.
		\item The gradient is scaled based on the step-length $\alpha$, applied to the starting model and an updated model is obtained $\mathbf{m}_{(k+1)}$.
		\item The process is iteratively repeated from Step \ref{itm:algo_fwi_2} until the convergence criterion is satisfied.
	\end{enumerate}
\label{algo:fwi}
\end{algorithm}

\subsection{Deep Neural Networks for FWI}
Neural Networks (NN) are a subset of tools in machine learning which when applied to inverse problems can approximate the non-linear functional of the inverse problem $F^{-1}:D\rightarrow M$. That is, using a NN, a non-linear mapping can be learned to minimize
\begin{equation}
\left||\mathbf{m} - g_{\theta}(\mathbf{d})\right||^2 ,
\end{equation}
where $\theta$ the large data set of pairs $(\mathbf{m}, \mathbf{d})$ used for the learning process \cite{Lucas2018}. 

The most elementary component in a NN is a neuron. This receives excitatory input and sums the result to produce an output or activation, representing a neuron's action potential which is transmitted along its axon \cite{Raschka2017}. For a given artificial neuron, consider $n$ inputs with signals $m$ and weights $w$. The output $d$ of the $k^{\text{th}}$ neuron from all input signals is given by:
\begin{equation}
d_k=\sigma\left( b+\sum_{j=0}^{m} w_{kj}m_{j} \right) ,
\end{equation}
where $\sigma$ is the activation function and $b$ is a bias term enabling the activation functions to shift about the origin.
\begin{figure*}[b]
	\centering
	\includegraphics[width=0.8\textwidth, trim = {2.7cm 5.5cm 4.6cm 2.3cm}, clip]{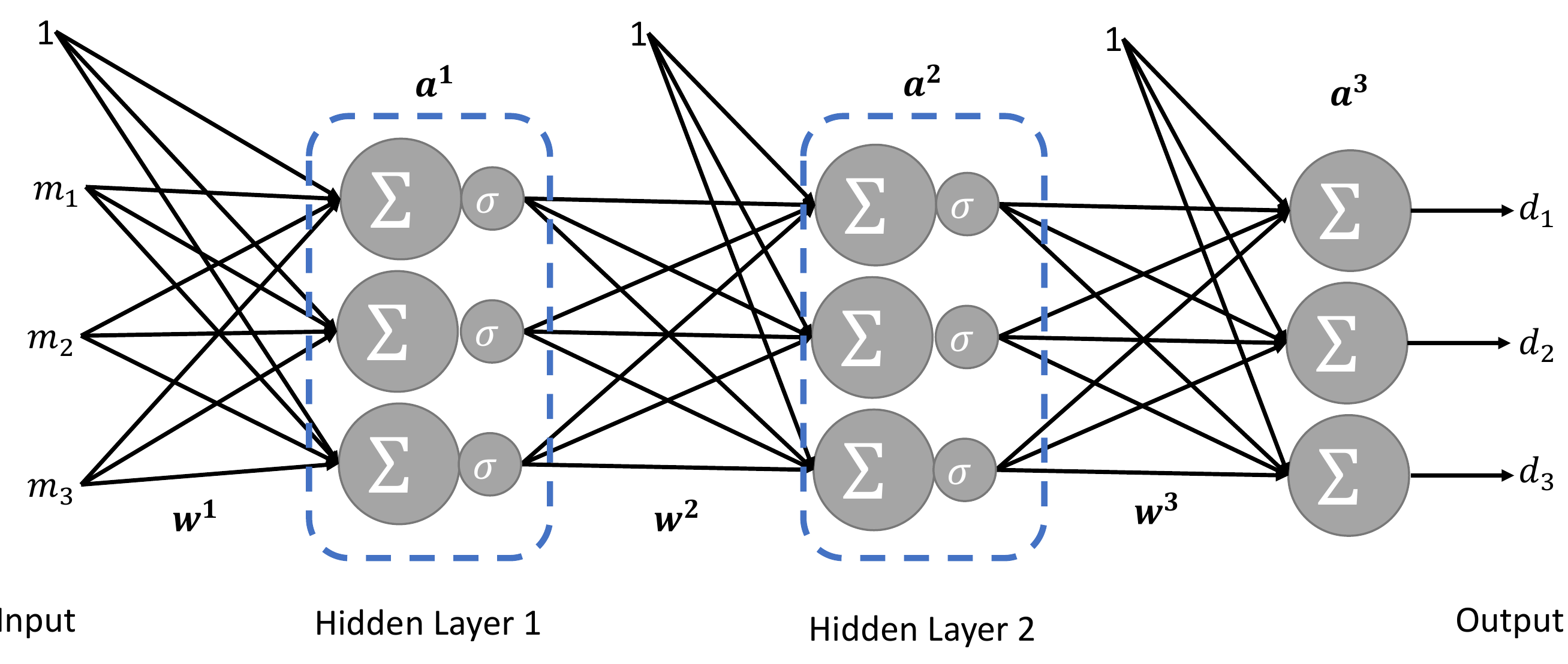}
	\caption{An example of a fully connected NN with 2 hidden layers. All weights $w$ and bias $b$ are learned during the training phase. The $1$'s connected to each hidden layer represents bias nodes which help the NN learn patterns by allowing the output of an activation function to be shifted. Adapted from \cite{Lucas2018}.}
	\label{fig:mlp_2_layer}
\end{figure*}
When multiple neurons are combined together, they form a NN. The architecture of a NN refers to the number of neurons, their arrangement and their connectivity \cite{Sima2003}. The initial layer of  nodes $\mathbf{m}$ are referred to as the Input Layer. These are connected to a sequence of hidden layers of neurons. The final layer of the neurons is not a hidden layer and is referred to as the Output Layer. Communication proceeds layer by layer from the input layer via the hidden layers up to the output layer. If a NN has two or more hidden layers, it is called a DNN. Figure \ref{fig:mlp_2_layer} shows a NN consisting of 2 hidden layers. The output of the unit in each layer is the result of the weighted sum of the input units, followed by a non-linear element-wise function. The weights between each units are learned as a result of a training procedure. 

When training a DNN, the forward propagation through the hidden layers from input $\mathbf{m}$ to output $\mathbf{d}$ needs to be measured for its misfit. The most commonly used cost function is the Sum of Squared Errors (SSE), defined as:
\begin{equation}
\emph{J}(\mathbf{m}) = \frac{1}{2}\sum_{i=1}^{J}\left( \mathbf{m} - g_{\theta}(\mathbf{d}^{(i)}) \right)^{2},
\end{equation}
where $\mathbf{d}$ is the labelled true dataset, $\mathbf{d}^{(i)}$ is the output from the $i^{\text{th}}$ forward pass through the network and the summation is across all neurons in the network. The objective is to minimize the function $\emph{J}$ with respect to the weights $w$ of the neurons in the NN. Employing the Chain Rule and after a series of recursive formulations, the error signals for all neurons in the network can be recursively calculated throughout the network and the derivative of the cost function with respect to all the weights $w$ can be calculated. Training of the DNN is then achieved via a Gradient Descent algorithm, referred to as back-propagation training algorithm \cite{rumelhart1985learning}. The reader is referred to \cite{Goodfellow-et-al-2016} and citations therein for a full mathematical formulation.

\subsection{Outline for solving FWI using DNN}
\begin{figure*}[h!]
	\centering
	\includegraphics[width=0.95\textwidth, trim = {4cm 1cm 1.5cm 2.5cm}, clip]{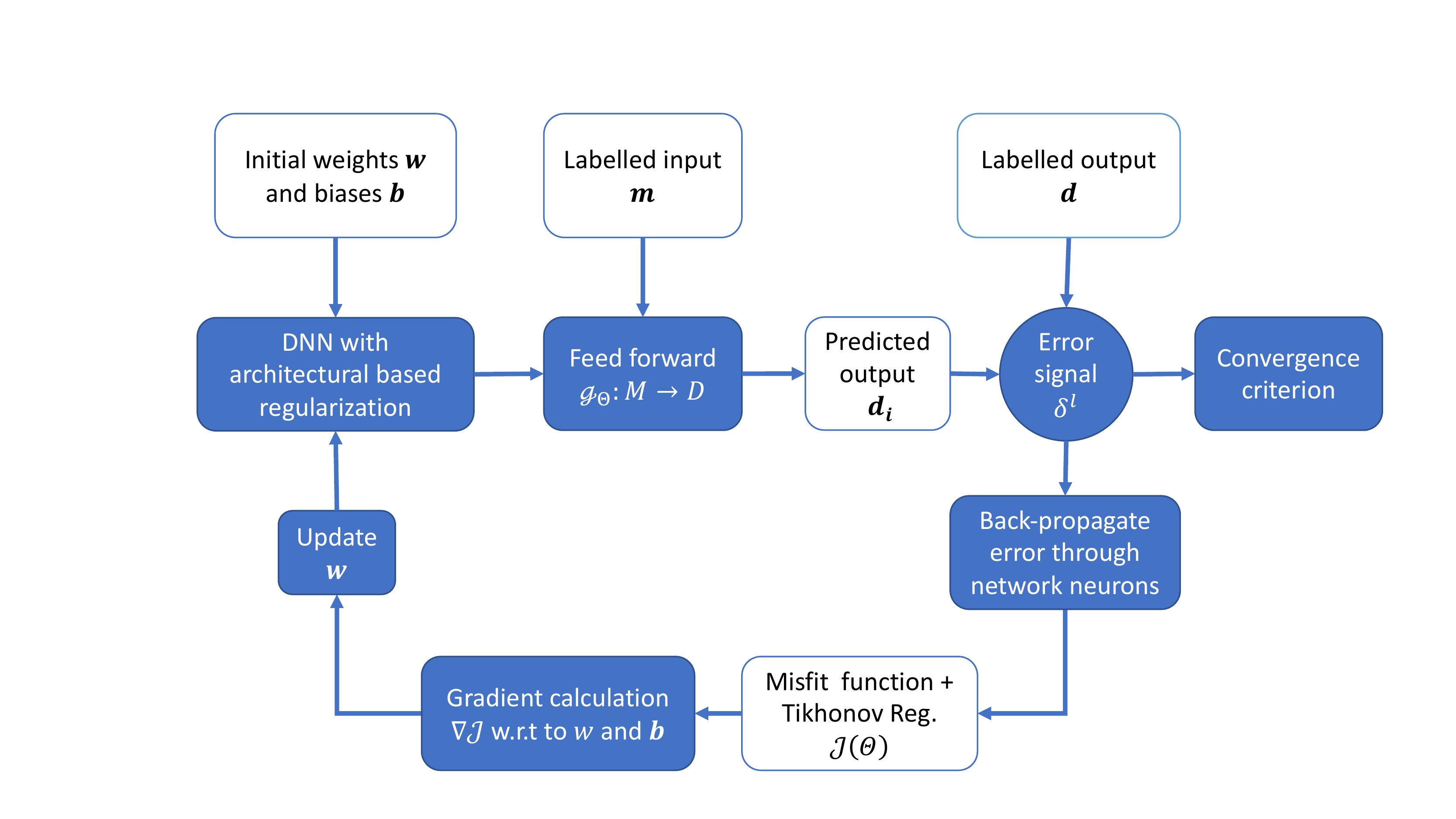}
	\caption{Schematic of a FWI workflow solved as learned optimisation process.}
	\label{fig:algo_schem_dnn}
\end{figure*}

Algorithm for training of a DNN for FWI is given in Algorithm \ref{algo:dnn} and a schematic is given in Figure \ref{fig:algo_schem_dnn}.
\begin{algorithm}
	\caption{FWI as a DNN problem}
	\renewcommand{\labelenumi}{(\Roman{enumi})}
	\begin{enumerate}[noitemsep,nolistsep]
		\item Setup a deep architecture for the NN.
		\item Initialise the set of weights $w^l$  and biases $b^l$.
		\item Forward propagate through the network connections to calculate input sums and activation function for all neurons and layers.
		\item Calculate the error signal for the final layer $\delta^L$ by choosing an appropriate differentiable activation function.
		\item Back-propagate the errors $(\delta^l)$ for all neurons in layer $l$.
		\item Differentiate the cost function with respect to biases $\left(\frac{\partial\emph{J}}{\partial b^l} \right) $.
		\item Differentiate the cost function with respect to weights $\left( \frac{\partial\emph{J}}{\partial w^l}\right)$.
		\item Update weights $w^l$ via gradient descent.
		\item Recursively repeat from Step 3 until the desired convergence criterion is met.
	\end{enumerate}
	\label{algo:dnn}
\end{algorithm}

\section{Numerical example}
\subsection{Experiment setup}
The hypothesis we would like to prove is as follows:

"Given a seismic trace in time domain, invert for the seismic velocity ($v_p$) via a DNN which transforms the input data into pseudo-spectral domain and learns to invert for a velocity estimate"

\subsection{Training data}\label{sec:training_data}
\begin{figure*}[h!]
	\centering
	\includegraphics[width=0.7\textwidth]{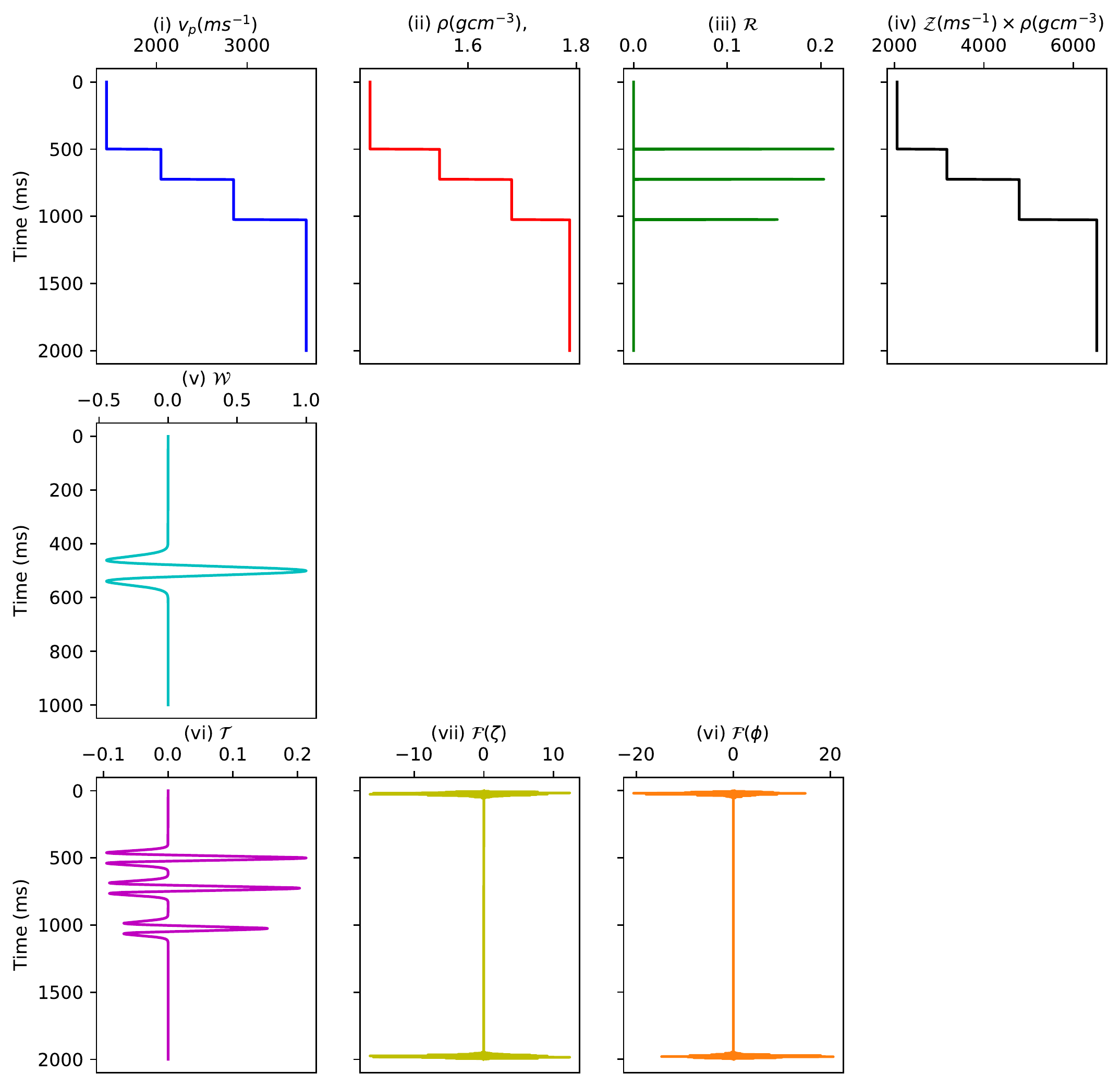}
	\caption{Workflow for creating a pseudo-spectral synthetic trace. This was repeated 500,000 times with random parameters generated within the pre-defined limits stated in Section \ref{sec:training_data} to create the learning dataset.}
	\label{fig:construction_synthetic}
\end{figure*}

Learning of the inversion from time to pseudo-spectral domain requires a training dataset which maps time to Fourier components of magnitude and phase, and their respective velocity profile. For our numeric example, 500,000 randomly generated mappings from time $(\mathcal{T})$ to Fourier components $(\mathcal{F})$ for a 2000ms time window were created. The steps involved in creation of the synthetic are shown in Figure \ref{fig:construction_synthetic} for a sample velocity profile and the steps involved in creating the dataset are given as:
\begin{enumerate}[label=\roman*]
	\item Randomly create a $v_p$ velocity profile for a $2000ms$ time duration, with values ranging from $1400ms^{-1}$ to $4000ms^{-1}$. The lower bound of $1400ms^{-1}$ was selected as in normal off-shore seismic exploration conditions, the smallest observed velocity is that of the water which ranges from $1450ms^{-1}$ to $1460ms^{-1}$ \cite{1560water}. The upper bound of $4000ms^{-1}$ was selected as this is the upper limit of velocity in porous and saturated sandstones \cite{lee1996seismic} and the assumption is made that limestones, carbonates and salt deposits are not present in the subsurface model being inverted as these have velocity ranges in excess of $4000ms^{-1}$.
	\item Calculate the density $\rho$ based on Gardner’s equation \cite{Gardner1974} given by $\rho=\alpha v_p^\beta$ where $\alpha=0.31$ and $\beta=0.25$ are empirically derived constants that depend on the Geology.
	\item At each interface, calculate the Reflection Coefficient $\mathcal{R} = \frac{\rho_2 v_{p_2} - \rho_1 v_{p_1}}{\rho_2 v_{p_2} + \rho_1 v_{p_1}} $ where $\rho_{i}$ is density of medium $i$ and $v_{p_i}$ is the $p$-velocity in medium $i$
	\item For each medium, calculate the Acoustic Impedance $\mathcal{Z}_i=\rho_i v_{p_i}$
	\item Define a wavelet $\mathcal{W}$. This was selected to be a Ricker wavelet at 10\si{Hz} \cite{ricker1943}. The Ricker wavelet is a theoretical waveform that takes into account the effect of Newtonian viscosity and is representative of seismic waves propagating through visco-elastic homogeneous media \cite{wang2015frequencies}, thus making it ideal for this numerical simulation. The central frequency of 10\si{Hz} was chosen as a nominal value based on literature results to be representative of normal FWI conditions \cite{morgan2013next}. Beyond 10\si{Hz} would be considered to be super-high-resolution FWI \cite{mispel2019high}, which goes beyond the scope of this manuscript.
	\item The Reflection Coefficient and wavelet are convolved to produce the seismic trace $\mathcal{T}$
	\item Fourier coefficients for magnitude $\mathcal{F}(\zeta)$ and phase $\mathcal{F}(\phi)$ are derived based on the FFT.
\end{enumerate} 

\subsection{DNN architecture}\label{sec:DNN_arch}
\begin{figure*}[h]
	\centering
	\includegraphics[width=0.7\textwidth, trim = {2cm 0cm 2cm 0cm}, clip]{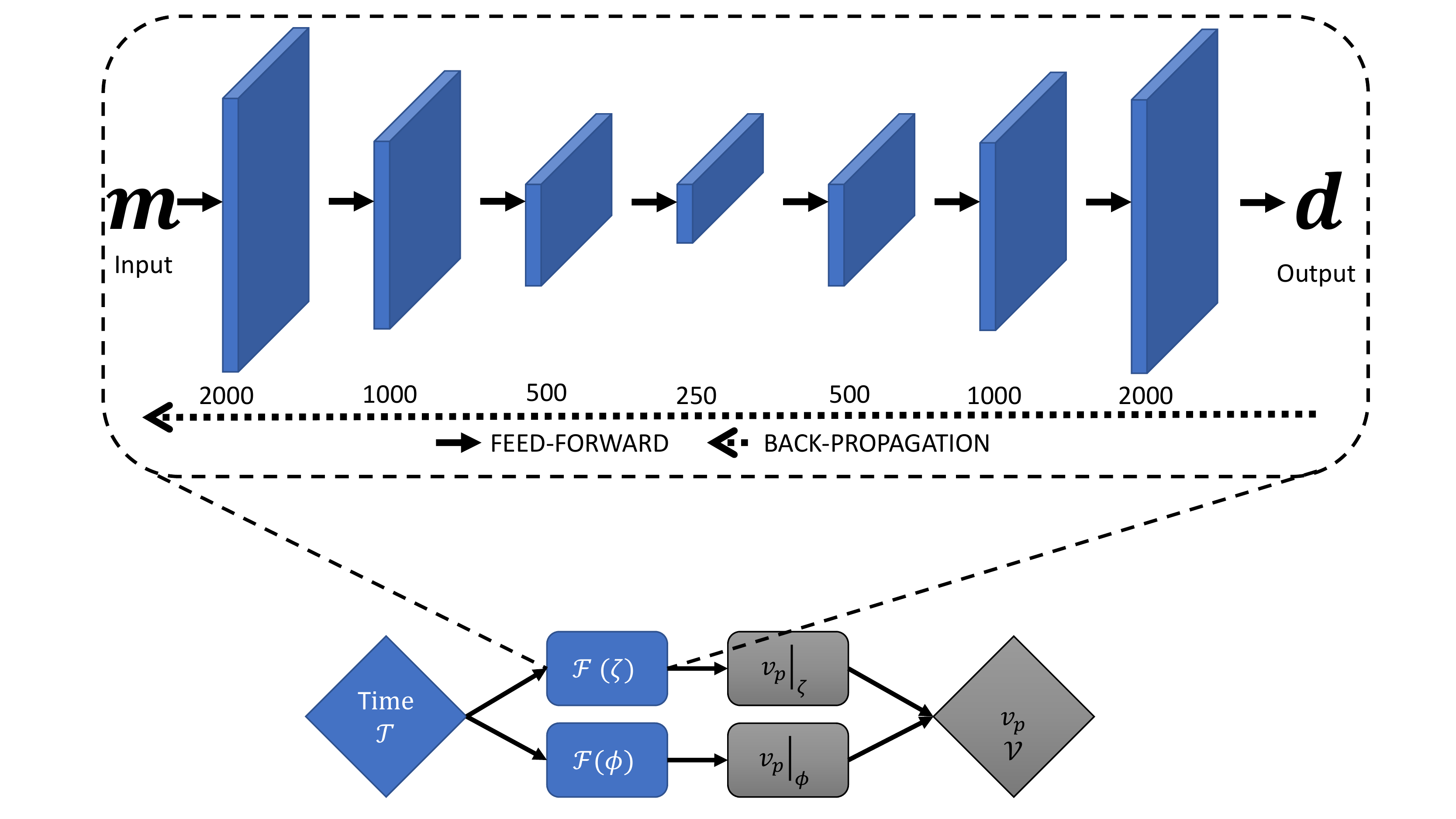}
	\caption{Pseudo-spectral FWI DNN architecture. The highlighted section indicates the set-up employed in each of the 5 modules. Each network has an hour-glass shape with layers of sizes 2000-1000-500-250-500-1000-2000 neuron which can be related to multi-scale FWI. The bottom section illustrates the DNN workflow where $\mathcal{T}$ is the input time domain, $\mathcal{V}$ is the output $v_p$ velocity and $\mathcal{F}$ is the Fourier domain, with magnitude $\zeta$ and phase $\phi$.}
	\label{fig:net_arch}
\end{figure*}

Figure \ref{fig:net_arch} illustrates the NN architecture used to first invert for the Fourier coefficients from the time domain and then invert for velocity profile. The complete workflow had 5 modules, with each module consisting of NN with 5 fully-connected hidden layers. The layer distributions consisted of an input layer of 2000 neurons, then a set of 5 hidden layers of sizes 1000, 500, 250, 500, 1000 neurons, and an output layer of 2000 neurons. This hour-glass design can be considered representative of multi-scale FWI \cite{Bunks1995} since at each hidden layer, the NN learns an abstracted component of the data at a different scale. The network employed a sum of squared errors loss function, data batching, early stopping, $L_2$ - norm regularization updates and executed for 200 epochs. A Rectified linear unit or ReLU function given by $f(x)=\max(0,x)$ was used as an activation function. This is a non-linear function which allows for back-propagation of errors. When employed on a network of neurons, the negative component of the function is converted to zero and the neuron is deactivated, thus introducing sparsity within the network and making it efficient and easy for computation. The output from each parallel thread in the flow is fed into another neural network which learns the optimal way of combining the outputs. In total, the DNN had 25 hidden layers. The learning or back-propagation for each network was optimized via an ADAM optimizer \cite{adam}, which is a stochastic gradient descent based algorithm for first-order gradient-based optimisation which employs on adaptive estimates of lower-order moments. The DNN was implemented in Python 3.7, using Keras 2.2.4 \cite{keras} and TensorFlow 1.13.1 \cite{abadi2016tensorflow} backend.

\subsection{Numerical results}
\begin{figure*}
	\centering
	\includegraphics[width=0.9\textwidth]{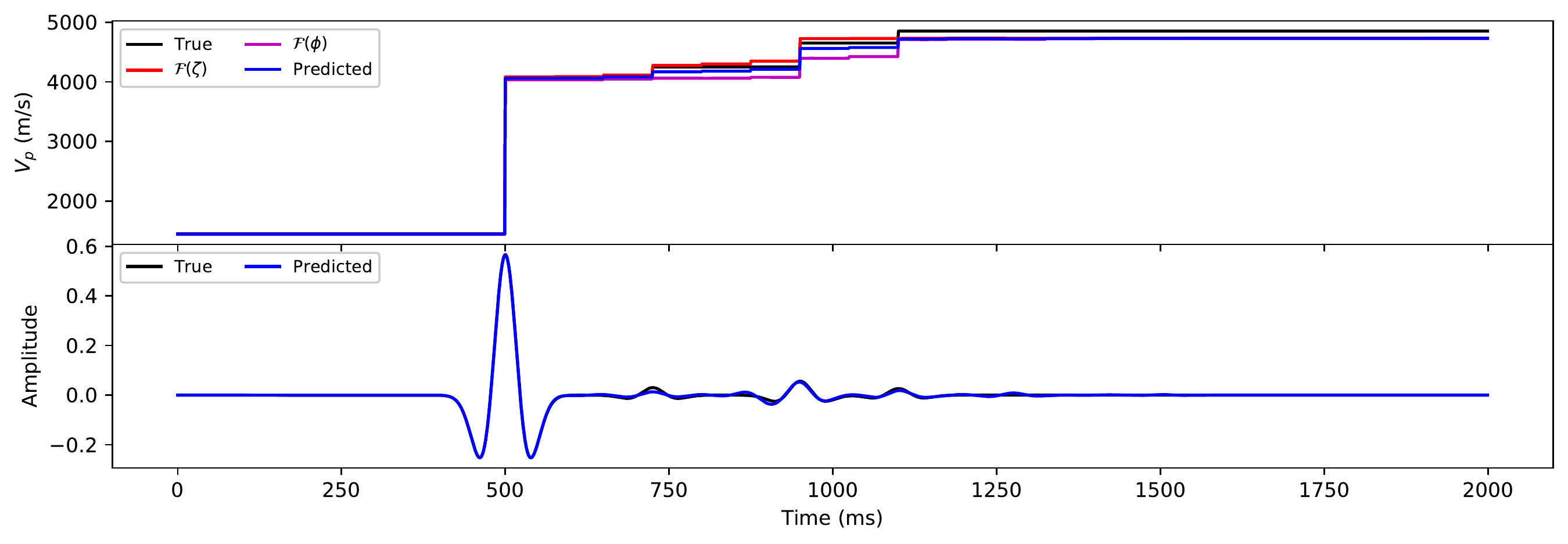}
	\includegraphics[width=0.9\textwidth]{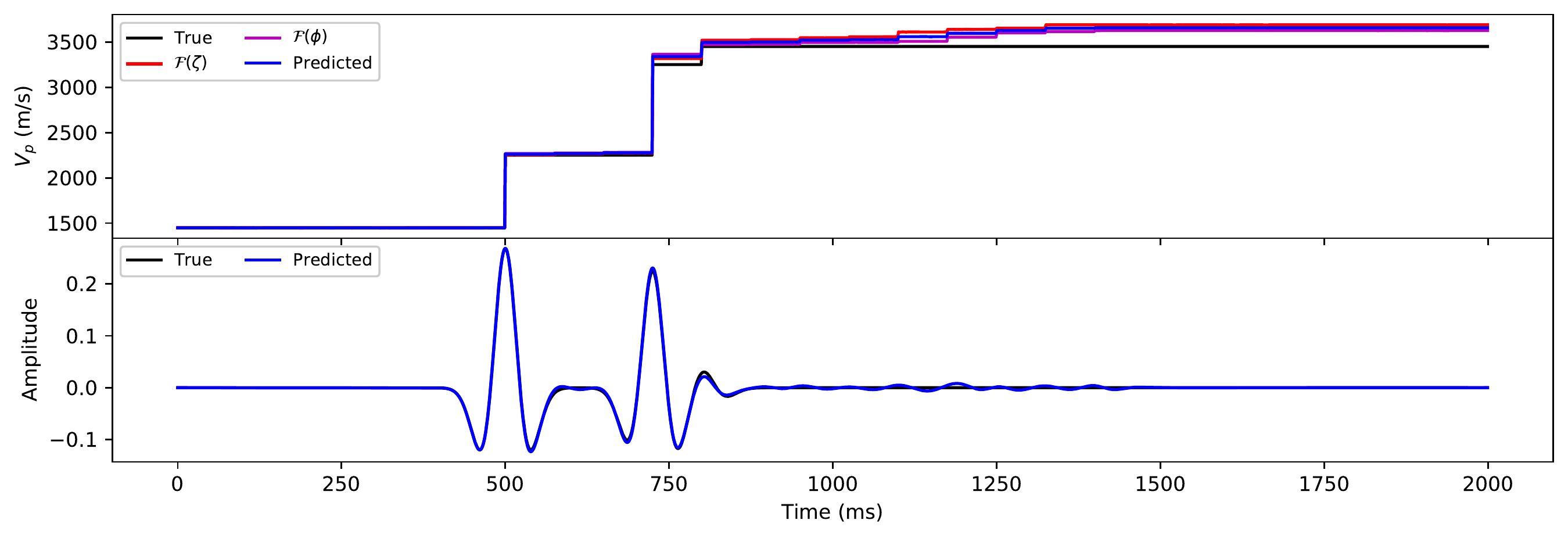}
	\includegraphics[width=0.9\textwidth]{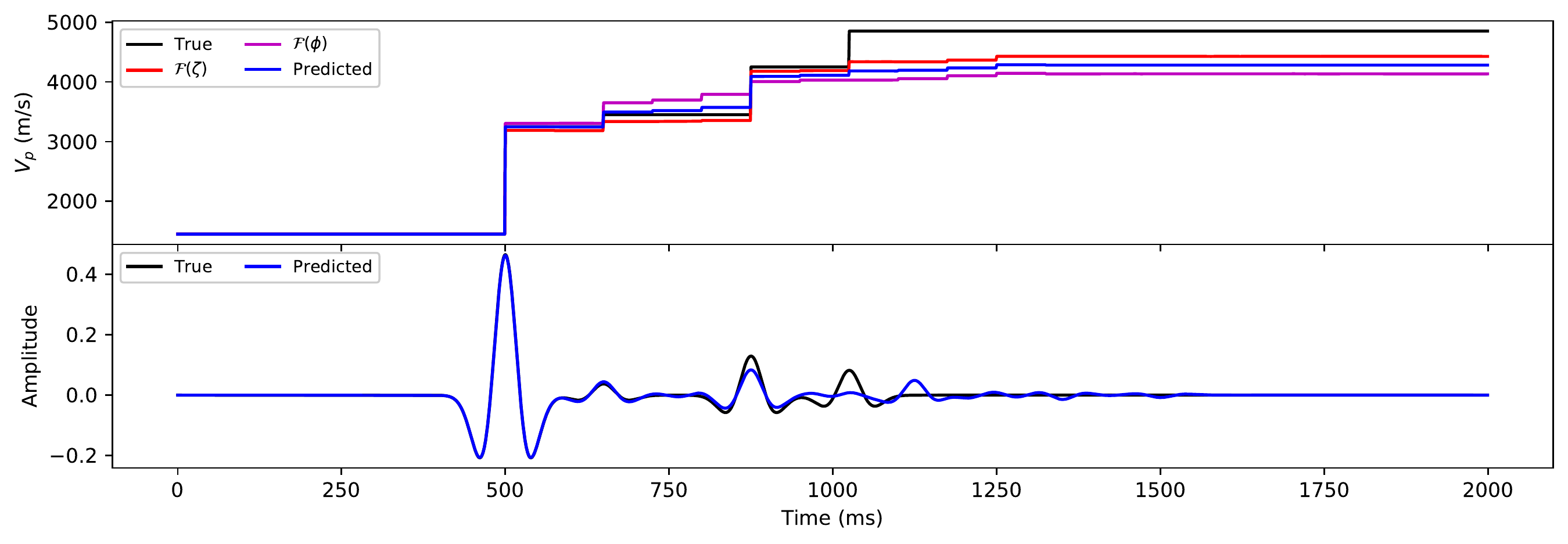}
	\includegraphics[width=0.9\textwidth]{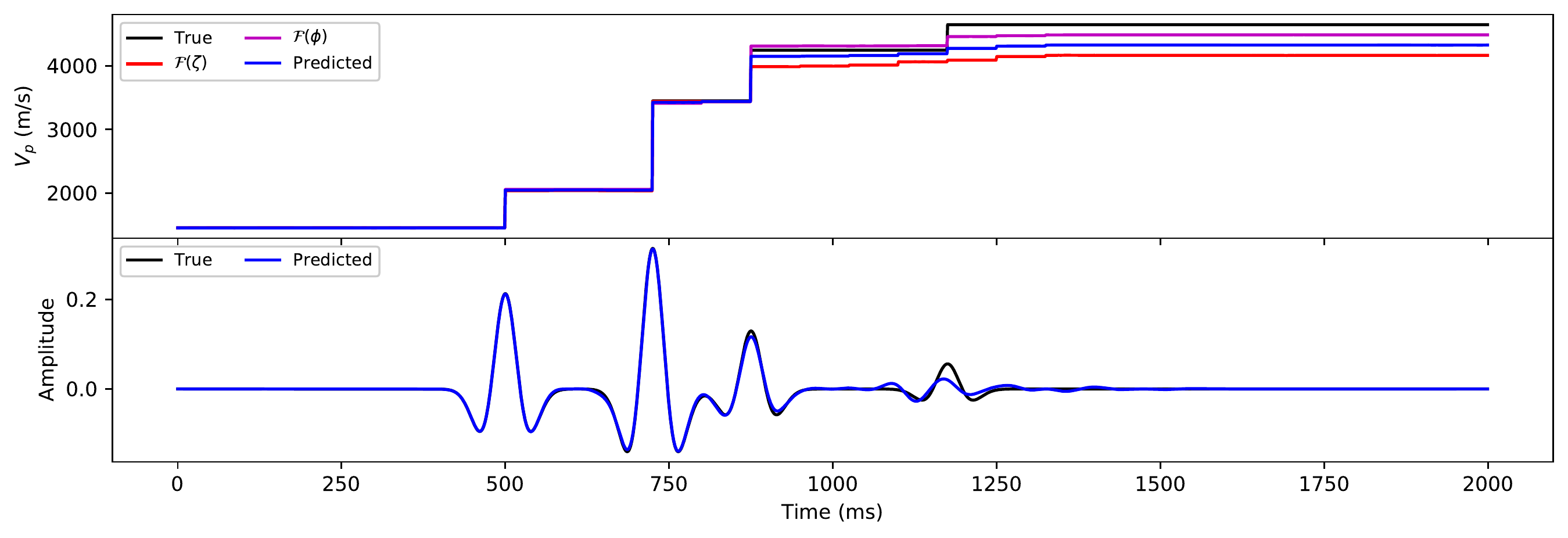}
	\caption{Four different predictions obtained from learned weights of the DNN on unseen data. The top panels are the velocity profile reconstructions from the two NN architecture branches ($\mathcal{F}(\zeta)$ and $\mathcal{F}(\phi)$) and the combined result. Bottom panels are the observed and inverted waveforms.}
	\label{fig:1d_results}	
\end{figure*}
\begin{figure*}

	\centering
	\includegraphics[width=0.9\textwidth,  trim = {0.4cm 11.9cm 0.4cm 0.1cm}, clip]{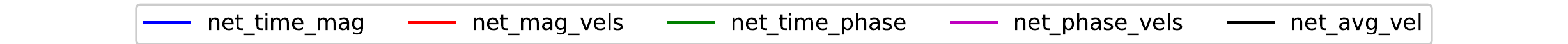}
	\hfill
	\subfigure[Training dataset MSE over the different epochs per DNN component. Overall performance is decreasing per epoch, indicating that the DNN is learning to invert.]
	{
		\label{fig:dnn_perf_metric_train}
		\includegraphics[width=0.4\textwidth,  trim = {0.3cm 0.2cm 20cm 0cm}, clip]{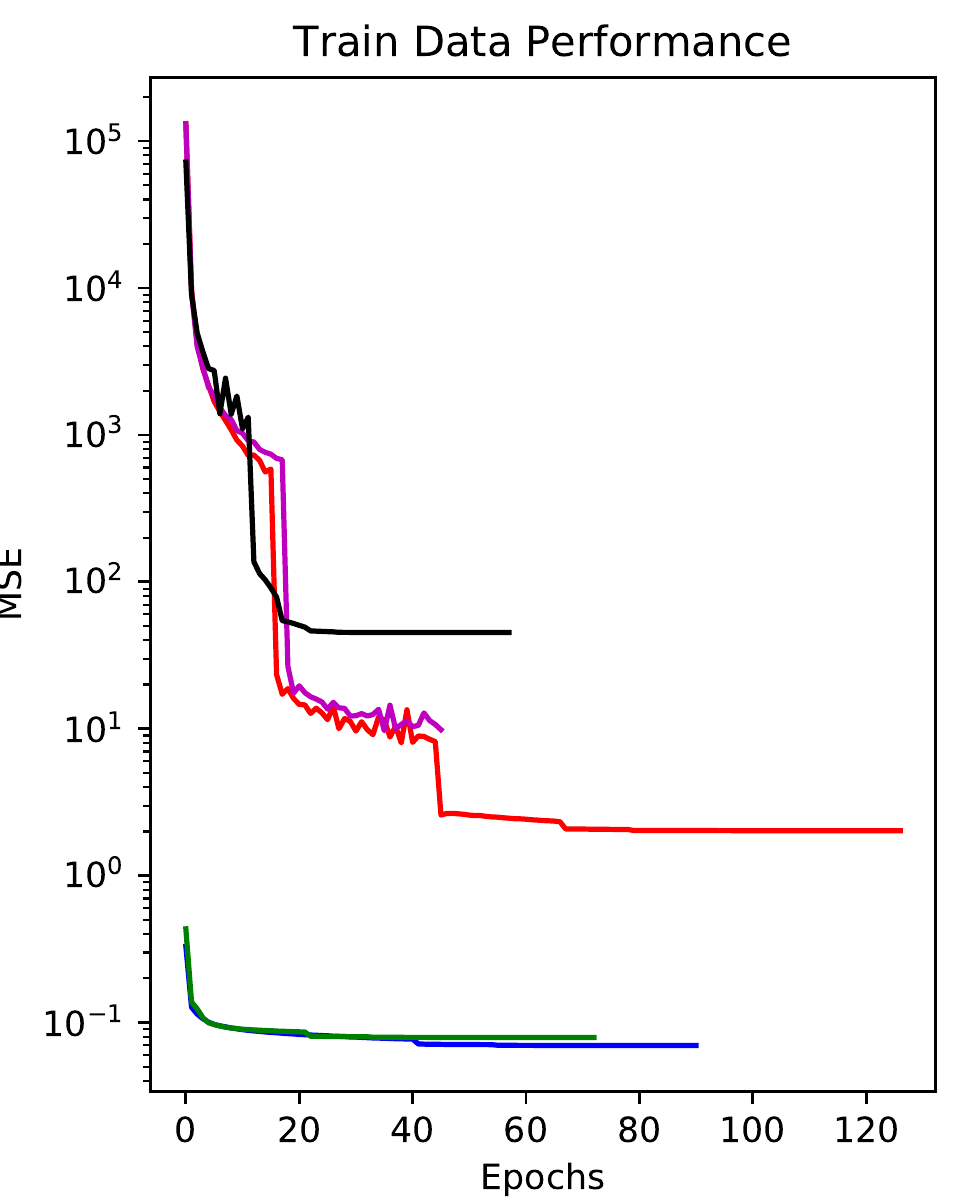}
		}
	\hspace{10pt}
	\subfigure[Test dataset MSE over the different epochs per DNN component.]
	{
		\label{fig:dnn_perf_metric_test}
		\includegraphics[width=0.4\textwidth,  trim = {10.2cm 0.2cm 10.1cm 0cm}, clip]{2_multilayer_7_xjenza_DNN_perf_train_test.eps}
		}
	\hfill
	\subfigure[Learning Rate performance over the different epochs per DNN component.]
	{
		\label{fig:dnn_perf_lr}
		\includegraphics[width=0.4\textwidth,  trim = {20cm 0.2cm 0.15cm 0cm}, clip]{2_multilayer_7_xjenza_DNN_perf_train_test.eps}
	}
	\caption{DNN  performance metrics.}
	\label{fig:1d_results_perf}
\end{figure*}
Figure \ref{fig:1d_results} illustrates the application of DNN architecture in Section \ref{sec:DNN_arch} for a sample of unseen data and the respective reconstruction. Inspection of the first 750\si{ms} indicates that the DNN approach is able to reconstruct both the velocity and the waveform profile near perfectly, irrespective of the number of layers and the magnitude of the acoustic difference in this time range. Beyond 750\si{ms}, reconstructions start suffering from slight degradation. As illustrated in the velocity reconstruction of the middle figure, the inaccuracy is minimal and ranges $\pm$100\si{ms^{-1}}. However, this leads to perturbations in the reconstruction and does not allow for perfect matching. Further inspection suggests that the main source of error is due to the magnitude component of the network (red). To improve this error component, the network inverting for the magnitude component of the FFT would need to be trained and generalised further. 

Figure \ref{fig:1d_results_perf} shows the DNN metric performance over the different epochs per DNN component. Figure \ref{fig:dnn_perf_metric_train} and \ref{fig:dnn_perf_metric_test} illustrate the MSE performance for the training and testing dataset respectively. Considering the former, the plots indicate that the network is indeed learning since MSE is decreasing at each epoch. Comparing respective DNN components between the training and the testing dataset metrics, there is evidence of no under-fitting or over-fitting with the pseudo-spectral learning components of the DNN architecture (\texttt{net\_time\_mag}, \texttt{net\_mag\_vels}, \texttt{net\_time\_phase}, \texttt{net\_phase\_vel}) and there is indeed good-fit since training and testing MSE both decrease to a point of stability with a minimal difference between the two final MSE values. On the other hand, \texttt{net\_avg\_vel} component which is learning to average out the velocity from Fourier components indicates symptoms of an under-presented training dataset. Moreover, these MSE performance plots indicate that the technique might suffer from a $\emph{compounding error}$ issue. The two best performing components are the first layer of learning for the inversion, namely Time-to-FFT-Magnitude (\texttt{net\_time\_mag}) and Time-to-FFT-Phase (\texttt{net\_time\_phase}), as their MSE performance plateaus at $10^{-1}$. In the second phase of the inversion which converts respective FFT components to velocities (FFT-Magnitute-to-Velocity (\texttt{net\_mag\_vels}) and FFT-Phase-to-Velocity (\texttt{net\_phase\_vels})), the error plateaus are at $10^1$, which is two orders of magnitude greater. The final network component sits even higher on the scale at $10^2$. Both the train and the test dataset show drastic decreases in the MSE at different epoch levels. These can be attributed to the step-wise reductions in learning rate shown in Figure \ref{fig:dnn_perf_lr}. This varying learning rate allows the network to move to a deeper optimisation level and approach a more global minima for the optimisation problem. 

\section{Conclusions}
In this manuscript we presented the investigation of direct modelling for seismic waveforms using a DNN which first converts data to pseudo-spectral domain and inverts for velocity profiles. Experimental results demonstrated that the use of synthetically generated data to train a DNN proves to be a viable technique to learn how to invert via pseudo-spectral data. Although inversion was successfully achieved in the numerical examples presented, one branch of the DNN architecture was lacking in inversion performance and was resulting in a compounding error effect. To improve the overall performance of the technique, data augmentation will be considered as potentially 500,000 random traces are not sufficient to train the magnitude component of the Fourier transform for the network to achieve a desirable performance, and fine-tuning of the NN architecture in the form of in-between layer regularization, neuron drop-out during epoch training and convolutional layers have yet to be investigated. Moreover, in the next stage, this technique will be used for the inversion of more interesting subsurface structures which have a geological relevance, evaluate image resolution when compared to standard FWI and also consider the case of a sequential input in the form of a Recurrent Neural Network, similar to the work of \cite{sun2019theory}, but via a pseudo-spectral approach.

\bibliography{references}
\end{document}